\documentclass[sigconf]{acmart}

\usepackage{amscd,amsfonts,amsbsy,rotating}
\usepackage{balance}
\usepackage{graphicx}
\usepackage{epsfig,epstopdf}
\usepackage{subfigure}
\usepackage{multirow}
\usepackage{booktabs}
\usepackage{color,xcolor}
\usepackage{url}
\usepackage{latexsym,bm}
\usepackage{enumitem,balance,mathtools}
\usepackage{wrapfig}
\usepackage{euscript}
\usepackage{algorithm}
\usepackage{algorithmic}
\usepackage{ifpdf}
\usepackage{diagbox}
\usepackage{caption}
\usepackage{makecell}
\usepackage{bm}
\usepackage{sidecap}

\usepackage{array}
\newcolumntype{N}{@{}m{0pt}@{}}

\usepackage{lipsum}

\theoremstyle{plain}

\usepackage{tikz}
\newcommand*{\circled}[1]{\lower.7ex\hbox{\tikz\draw (0pt, 0pt)%
		circle (.5em) node {\makebox[1em][c]{\small #1}};}}

\newcommand{\minisection}[1]{\vspace{5pt}\noindent\textbf{#1.}}

\AtBeginDocument{%
  \providecommand\BibTeX{{%
    \normalfont B\kern-0.5em{\scshape i\kern-0.25em b}\kern-0.8em\TeX}}}

\acmSubmissionID{rfp0122}

\copyrightyear{2020}
\acmYear{2020}
\setcopyright{acmcopyright}\acmConference[KDD '20]{Proceedings of the 26th ACM SIGKDD Conference on Knowledge Discovery and Data Mining}{August 23--27, 2020}{Virtual Event, CA, USA}
\acmBooktitle{Proceedings of the 26th ACM SIGKDD Conference on Knowledge Discovery and Data Mining (KDD '20), August 23--27, 2020, Virtual Event, CA, USA}
\acmPrice{15.00}
\acmDOI{10.1145/3394486.3403050}
\acmISBN{978-1-4503-7998-4/20/08}

\begin{document}
\fancyhead{}
\title{An Efficient Neighborhood-based Interaction Model for Recommendation on Heterogeneous Graph}

\author{
	Jiarui Jin$^{1}$, Jiarui Qin$^{1}$, Yuchen Fang$^{1}$, Kounianhua Du$^{1}$, Weinan Zhang$^{1}$, Yong Yu$^{1}$,\\ Zheng Zhang$^2$, Alexander J. Smola$^2$.}
\affiliation{$^1$Shanghai Jiao Tong University, $^2$Amazon Web Services}
\email{{jinjiarui97, qjr1996, arthur\_fyc, 774581965, wnzhang, yyu}@sjtu.edu.cn, zhaz@amazon.com, alex@smola.org}

\renewcommand{\shortauthors}{J. Jin, et al.}

\begin{abstract}
There is an influx of heterogeneous information network (HIN) based recommender systems in recent years since HIN is capable of characterizing complex graphs and contains rich semantics.
Although the existing approaches have achieved performance improvement, while practical, they still face the following problems.
On one hand, most existing HIN-based methods rely on explicit path reachability to leverage path-based semantic relatedness between users and items, \emph{e.g.}, metapath-based similarities. These methods are hard to use and integrate since path connections are sparse or noisy, and are often of different lengths.
On the other hand, other graph-based methods aim to learn effective heterogeneous network representations by compressing node together with its neighborhood information into single embedding before prediction. This weakly coupled manner in modeling overlooks the rich interactions among nodes, which introduces an early summarization issue.
In this paper, we propose an end-to-end Neighborhood-based Interaction Model for Recommendation (NIRec) to address above problems.
Specifically, we first analyze the significance of learning interactions in HINs and then propose a novel formulation to capture the interactive patterns between each pair of nodes through their metapath-guided neighborhoods.
Then, to explore complex interactions between metapaths and deal with the learning complexity on large-scale networks, we formulate interaction in a convolutional way and learn efficiently with fast Fourier transform. 
The extensive experiments on four different types of heterogeneous graphs demonstrate the performance gains of NIRec comparing with state-of-the-arts.
To the best of our knowledge, this is the first work providing an efficient neighborhood-based interaction model in the HIN-based recommendations.
\vspace{-2mm}
\end{abstract}



\settopmatter{printacmref=false, printfolios=false}


\maketitle

{\fontsize{8pt}{8pt} \selectfont
	\textbf{ACM Reference Format:}\\
	Jiarui Jin, Jiarui Qin, Yuchen Fang, Kounianhua Du, Weinan Zhang, Yong Yu, Zheng Zhang, Alexander J. Smola. 2020. An Efficient Neighborhood-based Interaction Model for Recommendation on Heterogeneous Graph. In \textit{Proceedings of the 26th ACM SIGKDD Conference on Knowledge Discovery and Data Mining (KDD '20), August 23--27, 2020, Virtual Event, CA, USA} ACM, New York, NY, USA, 10 pages.
	\url{https://doi.org/10.1145/3394486.3403050}}

\vspace{-2mm}
\section{Introduction} 

During the recent decade, the techniques of recommender systems have developed from pure collaborative filtering based on only user-item interactions \cite{koren2009matrix} to deep neural networks with various kinds of auxiliary data containing complex and useful information \cite{qu2018product,guo2017deepfm}.

Recently, the heterogeneous information network (HIN)\footnote{The terms ``heterogeneous information network'' and ``heterogeneous graph'' are used interchangeably in the related literature \cite{shi2018heterogeneous,wang2019heterogeneous}. In this paper, we mainly use ``heterogeneous information network'' (HIN).}, consisting of multiple types of nodes and/or links, has been leveraged as a powerful modeling method to fuse complex information and successfully applied to many recommender system tasks, which are called HIN-based recommendation methods \cite{shi2018heterogeneous,shi2016integrating}.
In Figure~\ref{fig:instance}, we present an instance of movie data characterized by an HIN.
We can easily see that the HIN contains multiple types of entities connected by different types of relations.
A variety of graph representation learning methods have been proposed on HINs to capture the rich semantic information, which roughly fall into the following two categories. 

One school is graph-based methodologies such as HetGNN \cite{zhang2019heterogeneous}, where deep neural network architectures such as GCN \cite{kipf2016semi} on homogeneous graphs are extended 
to enhance aggregating feature information of neighboring nodes on heterogeneous graphs.
However, these technologies usually compress the information of a node and its neighborhood into single embedding vector before making prediction \cite{fu2017hin2vec,shi2018heterogeneous}.
In this case, only two nodes and one edge are activated, yet other nodes and their connections are mixed and relayed, which introduces an \textbf{early summarization} issue \cite{qu2019end}.
The other school is metapath-based approaches.
A metapath means composite relation connecting two objects in network schema level. It has been adopted to capture the semantic information \cite{liu2018interactive,wang2019heterogeneous}.
Taking the movie data in Figure~\ref{fig:instance} as an example, the relations between user and item can be revealed by the metapath User-User-Movie (UUM) for the co-user relation and User-Director-Movie (UDM) for the co-director relation.
However, as \citet{shi2018heterogeneous} stated, metapath-based methods, heavily relying on explicit path reachability, may obtain bad performance when path connections are sparse or noisy.
Also, rich structural information of nodes outside metapaths, \emph{i.e.}, their neighborhoods, is omitted in these approaches.\par

Based on the above analysis, when designing HIN, state-of-the-art methods have not well solved, even may not be aware of, the following challenges faced by HIN-based recommendation, which we address in this paper:

\begin{figure}[t]
	\centering
	\includegraphics[width=0.45\textwidth]{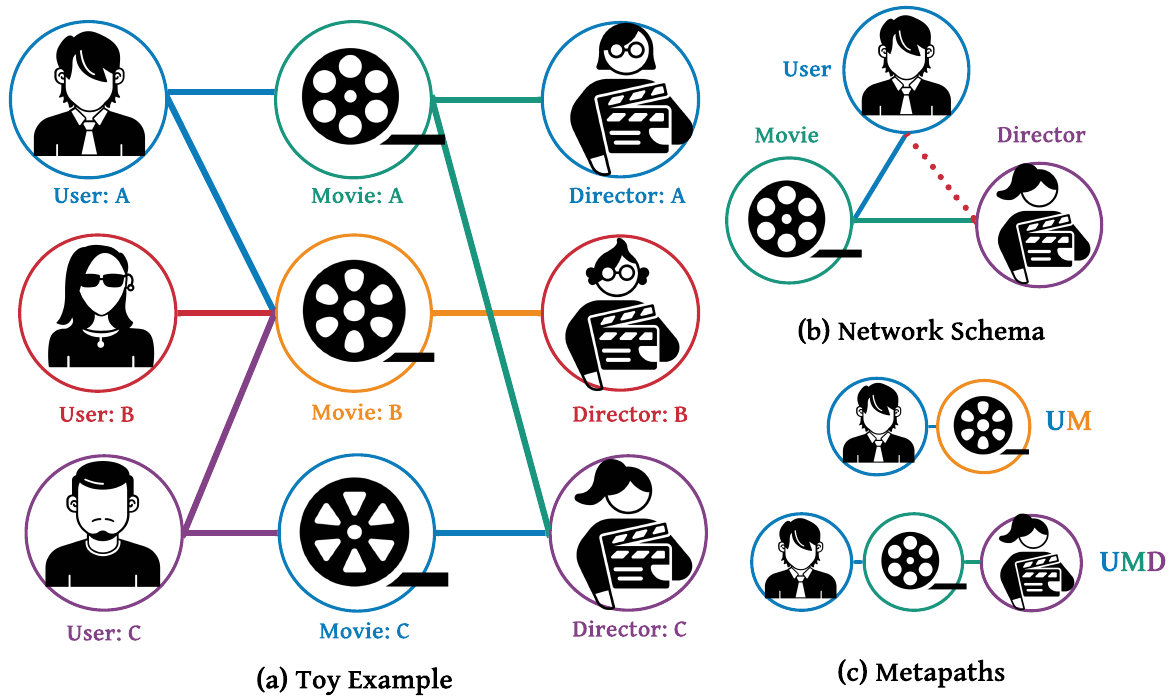}
	\vspace{-4mm}
	\caption{Heterogeneous information network and network schema. (a) movie $m_\text{A}$ and its metapath-guided neighbors (\emph{e.g}, $\mathcal{N}^2_\text{UMD}(u_\text{A}) = \{(u_\text{A}, m_\text{A}, d_\text{A}), (u_\text{A}, m_\text{A}, d_\text{C}), (u_\text{A}, m_\text{B}, d_\text{B})\}$) (b) three types of nodes (user, movie, director) and two types of relations/interactions (user-movie, movie-director) in solid line, potential interactions (user-director) in dashed line. (c) metapaths invloved (\emph{e.g.}, User-Movie (UM) and User-Movie-Director (UMD)).}
	\label{fig:instance}
	\vspace{-3mm}
\end{figure}

\begin{itemize}[topsep = 3pt,leftmargin =5pt]
	\item (\textbf{C1}) How to tackle the early summarization issue?
	Due to the complex structures and large scales of the graphs, it is hard to make predictions directly. 
	Hence, we consider the interactive local structures are valuable and helpful.
	For example, a system is recommending a user to a user $u_\text{A}$ based on an HIN as Figure~\ref{fig:instance} shows.
	When we consider a candidate user such as $u_\text{B}$, it is natural to consider their co-movies and co-directors; namely there may exist a relationship between $u_\text{A}$ and $u_\text{B}$ since they share the same movie $m_\text{B}$.
	Actually, this is an example of ``\textbf{AND}'' operation between users' neighborhoods.
	Also, it is easy to extend this into co-rating.
	In other words, the co-ratings between $u_\text{A}$ to $u_\text{B}$'s neighborhood and $u_\text{B}$ to $u_\text{A}$'s neighborhood indicate the similarity of users' preferences, which could be helpful for recommendation.
	We argue that this interactive local structures are hidden and not fully utilized in previous methods.

	\item (\textbf{C2}) How to design an end-to-end framework to capture and aggregate the interactive patterns between neighborhoods?
	A recent attempt \cite{liu2018interactive} focused on measuring the semantic proximity by interactive-paths connecting source and target nodes. However, it overlooked rich information hidden in these node neighborhoods.
	An HIN contains diverse semantic information reflected by metapaths \cite{sun2011pathsim}.	
	Also, there are usually various nodes in different types involved in one path.
	Different paths/nodes may contribute differently to the final performance. 
	Hence, besides a powerful interaction module, a well-designed aggregation module to distinguish the subtle difference of these paths/nodes and select some informative ones is required.
	
	
	\item (\textbf{C3}) How to learn the whole system efficiently?
	Learning interactive information on HINs is always time-consuming; especially when faced with paths in different types and lengths for metapath-based approaches \cite{liu2018interactive} and large-scale high-order information for graph-based approaches \cite{qu2019end}.
	A methodology to both efficiently and effectively learn the rich interactive information on HINs is always expected.
\end{itemize}

To tackle these challenges, we propose NIRec, a neighborhood-based interaction model for HIN-based recommendation.
First, we extend the definition of neighborhood in homogeneous graphs into metapath-guided neighborhood in heterogeneous graphs.
Next, we design a heterogeneous graph neural network architecture with two modules to aggregate feature information of sampled neighbors in previous step.
The first module, namely interaction module, constructs interactive neighborhood and captures latent information between ``\textbf{AND}'' operation.
The second module, namely aggregation module, mainly consists of two components: (i) node-level attention mechanism to measure the impacts of different nodes inside a metapath-guided neighborhood, and (ii) path-level attention mechanism to aggregate content embeddings of different neighborhoods.
Finally, we formulate interaction in a convolutional way and learn efficiently with fast Fourier transform (FFT). 
To summarize, the main contributions of our work are:\par

\begin{itemize}[topsep = 3pt,leftmargin =5pt]
	
	\item We formalize and address an important, but seldom exploited, early summarization issue on HIN.		
	\item We present an innovative convolutional neighborhood-based interaction model for recommendation on HINs, named NIRec, which is able to capture and aggregate rich interactive patterns in both node- and path-levels.
	\item We propose an efficient end-to-end learning algorithm incorported with fast Fourier transform (FFT).
	
\end{itemize}
We conduct extensive experiments on four public datasets. Our results demonstrate the superior performance of NIRec over state-of-the-art baselines.

\section{Related Work}
\label{sec:related-work}
\minisection{Heterogeneous Information Network based Recommendation}
As a newly emerging direction, heterogeneous information network (HIN) \cite{shi2016survey} can naturally characterize complex objects and rich relations in recommender systems.
There is a surge of works 
on learning representation in heterogeneous networks, \emph{e.g.} metapath2vec \cite{dong2017metapath2vec}, HetGNN \cite{zhang2019heterogeneous}, HIN2vec \cite{fu2017hin2vec}, eoe \cite{xu2017embedding}; and their applications, \emph{e.g.} relation inference \cite{sun2012will}, classification \cite{zhang2018deep}, clustering \cite{ren2014cluscite}, author identification \cite{chen2017task}.
Among them, HIN based recommendation has been increasingly attracting researchers' attention in both academic and industry fields.  
For instance,
\citet{feng2012incorporating} proposed to alleviate the cold start issue with heterogeneous information network contained in social tagged system.
Metapath-based methods were introduced into hybrid recommender system in \cite{yu2013recommendation}.
\citet{yu2014personalized} leveraged personalized recommendation framework via taking advantage of different types of entity relationships in heterogeneous informaiton network.
\citet{luo2014hete} proposed a collaborative filtering based social recommendation containing heterogeneous relations.
\citet{shi2015semantic} introduced weighted heterogeneous information network
In \cite{shi2016integrating}, the similarities of both users and items are evaluated unser dual regularization framework. 
Recently, \citet{hu2018leveraging} leveraged metapath-based context in top-N recommendation.
As stated in \cite{shi2018heterogeneous}, most existing HIN based recommendation methods rely on the path-based similarity, which may not fully mine latent features of users and items.
In this paper, we introduce metapath-guided neighborhood, and propose an innovative model to capture interactive patterns hidden in neighborhoods. \par

\minisection{Graph Representation}
Graph representation learning is mainly leveraged to learn latent, low dimensional representations of graph vertices, while preserving graph structure, \emph{e.g.}, topology structure and node content.
In general, graph representation algorithms can be categorized in two types.
One school is unsupervised graph representation algorithm, which aims at preserving graph structure for learning node representations \cite{perozzi2014deepwalk,grover2016node2vec,ribeiro2017struc2vec,tang2015line,wang2016structural}.
For instance, DeepWalk \cite{perozzi2014deepwalk} utilized random walks to generate node sequences and learn node representations.
Node2vec \cite{grover2016node2vec} further exploited a biased random walk strategy to capture more flexible contextual structures.
Struc2vec \cite{ribeiro2017struc2vec} constructed a multilayer graph to encode structural similarities and generate structural context for nodes.
LINE \cite{tang2015line} proposed an edge-sampling algorithm 
improving both the effectiveness and the efficiency of the inference.
SDNE \cite{wang2016structural} implied multiple layers of non-linear functions to capture highly non-linear network structure.
Another school is semi-supervised model \cite{huang2017label,kipf2016semi,velivckovic2017graph}, where there exist some labeled vertices for representation learning.
For example, LANE \cite{huang2017label} incorporated label information into the attributed network embedding while preserving their correlations.
GCN \cite{kipf2016semi} proposed a localized graph convolutions to improve the performance in a classification task.
GAT \cite{velivckovic2017graph} used self-attention network for information propagation, which leverages a multi-head attention mechanism.
GCN and GAT are popular architectures of the general graph networks and can be regarded as plug-in graph representation modules in heterogeneous graph, such as HetGNN \cite{zhang2019heterogeneous}, HAN \cite{wang2019heterogeneous} and HERec \cite{shi2018heterogeneous}.
In this work, our propose is not only to deliver graph representation learning techniques on heterogeneous graph, but also propose a novel efficient neighborhood-based interaction methods, which should be the first time to our knowledge.

\section{preliminary}
\label{sec:pre}	
In this section, we introduce the concept of the content-associated heterogeneous information network that will be used in the paper.

\vspace{-1mm}
\begin{definition}\textbf{Neighborhood-based Interaction-enhanced Recommendation.}
	The HIN-based recommendation task can be representetd as a tuple $\langle\mathcal{U}, \mathcal{I}, \mathcal{A}, \mathcal{R}\rangle$, where $\mathcal{U} = \{u_1, \ldots, u_p\}$ denotes the set of $p$ users; $\mathcal{I} = \{i_1, \ldots, i_q\}$ means the set of $q$ items; $a \in \mathcal{A}$ denotes the attributes associated with objects, and $r \in \mathcal{R}$ presents the interaction behaviors between different types of objects.
	The purpose of recommendation is to predict the interaction $r(u_s, i_t)$, \emph{i.e.}, click-through rate or link, between two objects, namely source user $u_s$ and target item $i_t$.
	In our setting, interactions between source and target neighbors are leveraged to enhance the performance.
\end{definition}
\vspace{-1mm}

To clarify the definition of the term ``neighborhood'', we model our recommendation task in the setting of \textbf{Heterogeneous Information Network} (HIN).
An HIN is defined as a graph $\mathcal{G}=(\mathcal{V}, \mathcal{E})$, which consists of more than one node type or link type.
In HINs, \textbf{network schema} is proposed to present the meta structure of a network, including the object types and their interaction relations.\par

Figure~\ref{fig:instance}(a) shows a toy example of HINs and the corresponding network schema is presented in Figure~\ref{fig:instance}(b).
We can see that the HIN consists of multiple types of objects and rich interaction relations.
The set of objects, \emph{i.e.}, $\mathcal{U}$ (user), $\mathcal{I}$ (movie) and $\mathcal{A}$ (director); interaction relations, \emph{i.e.}, $\mathcal{R}$ (relation), constitute $\mathcal{V}$ and $\mathcal{E}$ in the HIN, respectively. 
The interactions here can be explained as preferences between different types of nodes, \emph{e.g.}, $r(u_\text{A}, d_\text{C})$; and similarities between the same type of nodes, \emph{e.g.}, $r(u_\text{A}, u_\text{C})$.
In this case, user-item rating history, user-director preference, and movie-director knowledge constitute the interaction set $\mathcal{R}$.
When predicting interaction $r(u_\text{A}, m_\text{C})$ between source user $u_\text{A}$ and target movie $m_\text{C}$, we consider existing interactions between their neighborhood, namely $r(m_\text{A}, d_\text{C})$, $r(m_\text{B}, u_\text{C})$, to help the final recommendation.\par

In order to capture the structural and semantic relation, the metapath \cite{sun2011pathsim} is proposed as a relation sequence connecting two objects.
As illustrated in Figure~\ref{fig:instance}(c), the User-Movie-Director (UMD) metapath indicates that users favor movies, and that these movies are guided by some directors.
Based on metapaths, we introduce metapath-guided neighborhoods as follows.

\begin{definition}\textbf{Metapath-guided Neighborhood.}
	\label{def:neighbor}
	Given an object $o$ and a metapath $\rho$ (start from $o$) in an HIN, the metapath-guided neighborhood is defined as the set of all visited objects when the object $o$ walks along the given metapath $\rho$.
	In addition, we denote the neighbors of object $o$ after $i$-th steps sampling as $\mathcal{N}_\rho^i(o)$.
	Specifically, $\mathcal{N}_\rho^0(o)$ is $o$ itself.
	For convenience, let $\mathcal{N}_\rho(o)$ denote $\mathcal{N}^{I-1}_\rho(o)$, where $I$ means the length of metapath.
	It should be noted that different from similar concept proposed in \cite{wang2019heterogeneous} utilized to derive homegeneous neighbors on heterogeneous graph, metapath-guided neighborhood here preserves semantic content since it consists of heterogeneous information.
\end{definition}	

Taking Figure~\ref{fig:instance} as an instance, given the metapath ``User-Movie-Director (UMD)'' and a user $u_A$, we can get metapath-guided neighborhood as $\mathcal{N}_\rho^1(u_\text{A})$ = $\{(u_\text{A}, m_\text{A})$, $(u_\text{A}, m_\text{B})\}$, $\mathcal{N}_\rho(u_\text{A})$ = $\mathcal{N}_\rho^2(u_\text{A})$ = $\{(u_\text{A}, m_\text{A}, d_\text{A})$, $(u_\text{A}, m_\text{A}, d_\text{C})$, $(u_\text{B}, m_\text{B}, d_\text{B})\}$.

Many efforts have been made for HIN-based recommendations.
While, most of these works focus on leveraging graph neural networks to aggregating structural message, but overlook the effect of early summarization issue and rich interaction information.
Given the above preliminaries, we are ready to introduce our NIRec model as a solution.

\begin{figure*}[t]
	\centering
	\includegraphics[width=0.65\textwidth]{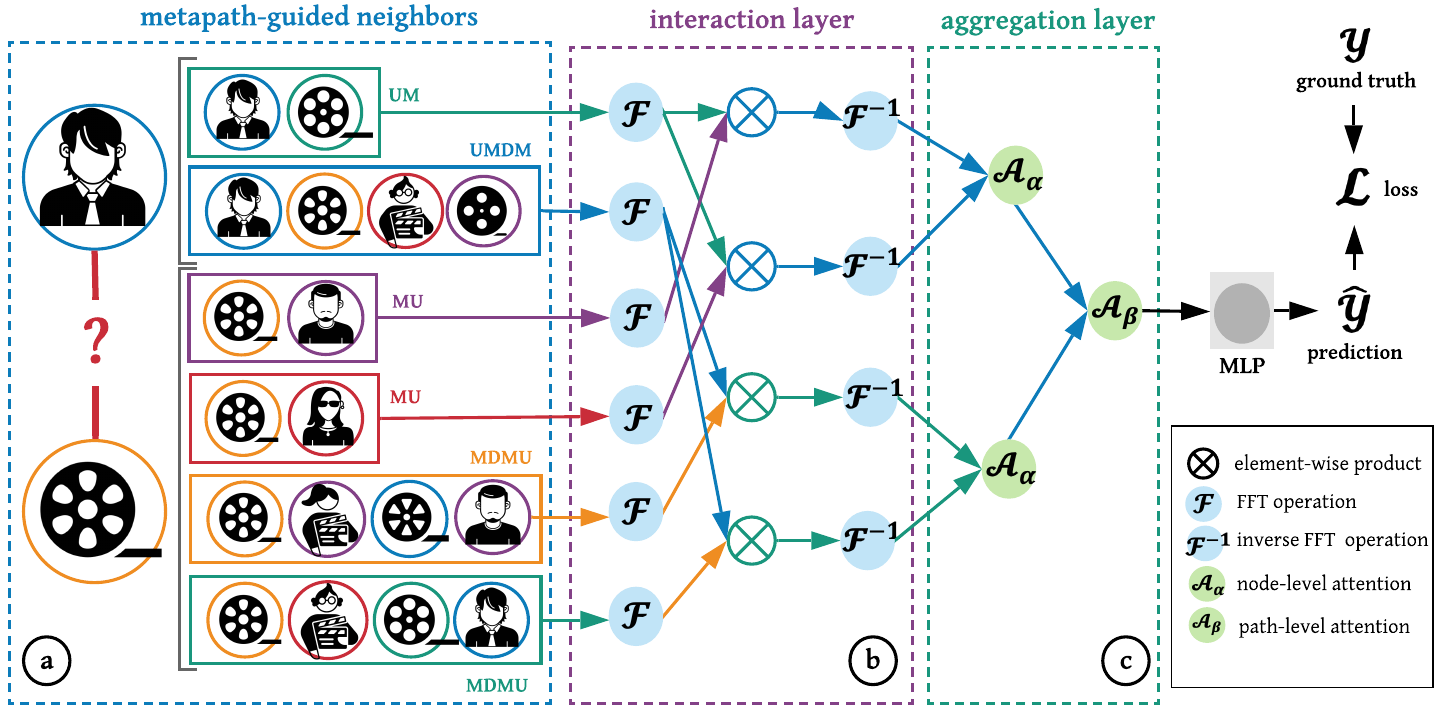}
	\vspace{-3mm}
	\caption {The overall architecture of NIRec: (a) it first samples metapath-guided neighborhood (Section~\ref{subsec:sample}); (b) next constructs interactive information via interaction layer (Section~\ref{subsec:interaction}); (c) finally combines rich information through aggregation layer (Section~\ref{subsec:aggregation}).}
	\label{fig:overview}
	\vspace{-4mm}
\end{figure*}

\section{Methodology}
\label{sec:method}
\subsection{Overview}
The basic idea of NIRec is to design a neighborhood-based interaction model to enhance the representations of objects.
With the help of HINs built on recommender systems, NIRec leverages metapaths to guide the selection of different-step and -type neighbors and design a heterogeneous interaction module to capture the abundant interaction message, and a heterogeneous aggregation module to obtain the rich embeddings of objects.
Moreover, we represent different types of metapaths with a uniform learning procedure and conduct the fast Fourier transform (FFT) as an efficient learning algorithm.  
\par

We provide the overall framework of NIRec in Figure~\ref{fig:overview}.
First, we utilize the multiple-object HIN containing $\langle$user, item, attribute, relation$\rangle$ as the input.
Second, we select metapath-guided neighbors for source and target nodes via neighbor samplings (Figure~\ref{fig:overview}(a)).
Third, we introduce the interactive convolutional operation to generate potential interaction information among their neighborhoods (Figure~\ref{fig:overview}(b)).
After that, we capture the key interactions and aggregate information via attention mechanism in both node and path level (Figure~\ref{fig:overview}(c)).
Finally, NIRec provides the final prediction.
We illustrate the architecture in detail in the following subsections.

\subsection{Neighborhood Sampling}
\label{subsec:sample}
To generate meaningful node sequences, the key technique is to design an effective random walk strategy that is able to capture the complex semantics reflected in HINs.
Hence, we propose to use the metapath guided random
walk method.
Giving a HIN $\mathcal{G} = \{\mathcal{V}, \mathcal{E}\}$ and a metapath $\rho: A_0, \cdots, A_i, \cdots, A_{I-1}$, where $A_i \in \mathcal{A}_i$ denotes $i$-th node guided by metapath $\rho$.
Note that we include user set $\mathcal{U}$ and item set $\mathcal{I}$ in attribute set $\mathcal{A}$ for convenience.
The walk path is generated according to the following distribution:
\begin{equation}
	\label{eqn:neighbor}
	P(n_{k+1}=x|n_k=v) = \left\{
	\begin{aligned}
		\frac{1}{|\mathcal{N}^1_\rho(v)|}, \ & \ (v, x) \in \mathcal{E} \ \text{and} \ v, x \in \mathcal{A}_k, \mathcal{A}_{k+1} \\
		0, \ & \ \text{otherwise}.
	\end{aligned}
	\right.
\end{equation}
where $n_k$ is the $k$-th node in the walk, $\mathcal{N}^1_\rho(v)$ means the first-order neighbor set for node $v$ guided metapath $\rho$.
A walk will follow the pattern of the metapath repetitively until it reaches the predefined length.
It is worth noting that, according to \emph{Definition~\ref{def:neighbor}}, there is no need to sample a complete metapath from source to target node.

\begin{figure}[b]
	\centering
	\vspace{-5mm}
	\includegraphics[width=0.35\textwidth]{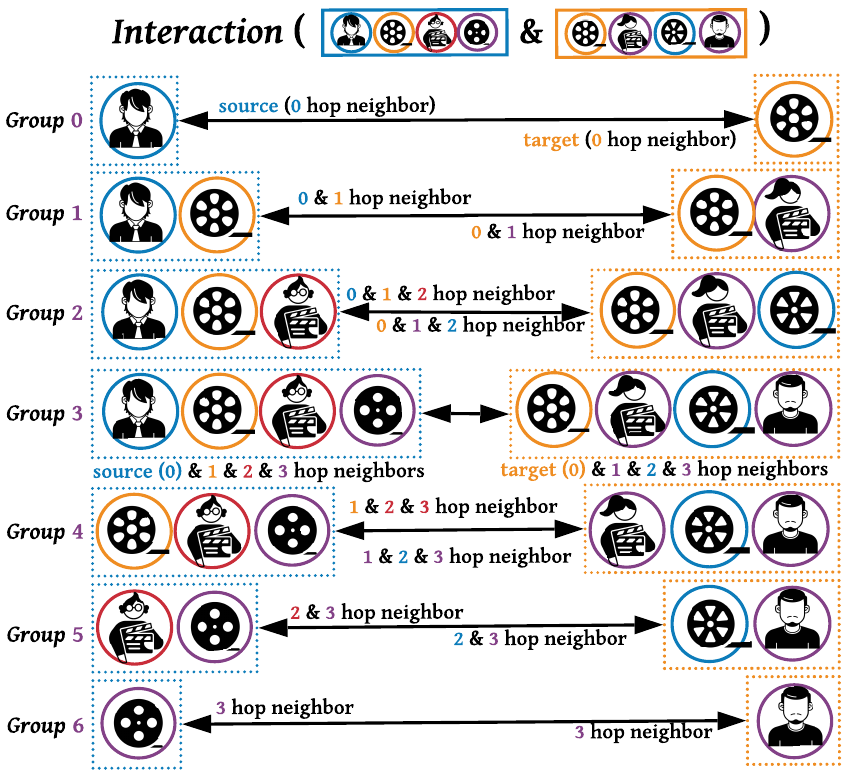}
	\vspace{-3mm}
	\caption {An illustrated example for the motivation in interaction module design. Neighborhood is grouped according to the distance to the source/target node. Interaction is only employed between corresponding neighborhoods.}
	\label{fig:mov}
	\vspace{-3mm}
\end{figure}

\subsection{Interaction Module}
\label{subsec:interaction}
In previous HIN-based recommendations, most approaches leverage graph representation techniques to find key nodes or metapaths \cite{wang2019heterogeneous} and capture the complex structure \cite{zhang2019heterogeneous}.
To further mine interaction information and deal with the early summarization issue, we propose an interaction module based on metapath-guided neighbors.\par

Due to the heterogeneity of nodes, different types of nodes have different feature spaces.
Hence, for each type of nodes (\emph{e.g.}, node with type $\phi_i$), we design the type-specific transformation matrix $\mathbf{M}_{\phi_i}$ to project the features of different types of nodes into a unified feature space.
The project process can be shown as follows:
\begin{equation} 
	\label{projection}
	e'_i = \mathbf{M}_{\phi_i} \cdot e_i,
\end{equation}
where $e_i$ and $e'_i$ are the original and projected features of node $i$, respectively.
By type-specific projection operation, our model is able to handle arbitrary types of nodes.\par

Considering that neighbors in different distances to the source/target node usually contribute differently to the final prediction, we divide the sampled metapath-guided neighborhood into several inner-distance and outer-distance neighbor groups.
As illustrated in Figure~\ref{fig:mov}, when we set distance as 1, we can get inner-distance \emph{Group} 1 and outer-distance \emph{Group} 5.
In a similar way, we can obtain $2I-1$ groups, where $I$ denotes the metapath length.
We argue that interactions should only be employed in corresponding groups.
In order to perform interaction in each neighbor group, we need to face two situations.
If there is only one node in \emph{Group}, we adopt element-wise product (``AND'') operation to measure their similarity or co-ratings, \emph{e.g.}, $r(u_\text{A}, m_\text{B})$ in \emph{Group} 0 case.
When there is more than one node, we first do interaction by production and then aggregate by summarization, \emph{e.g.}, $r(u_\text{A}, d_\text{C}) + r(m_\text{B}, m_\text{B})$ in \emph{Group} 1 case.\par

\begin{figure}[b]
	\centering
	\includegraphics[width=0.35\textwidth]{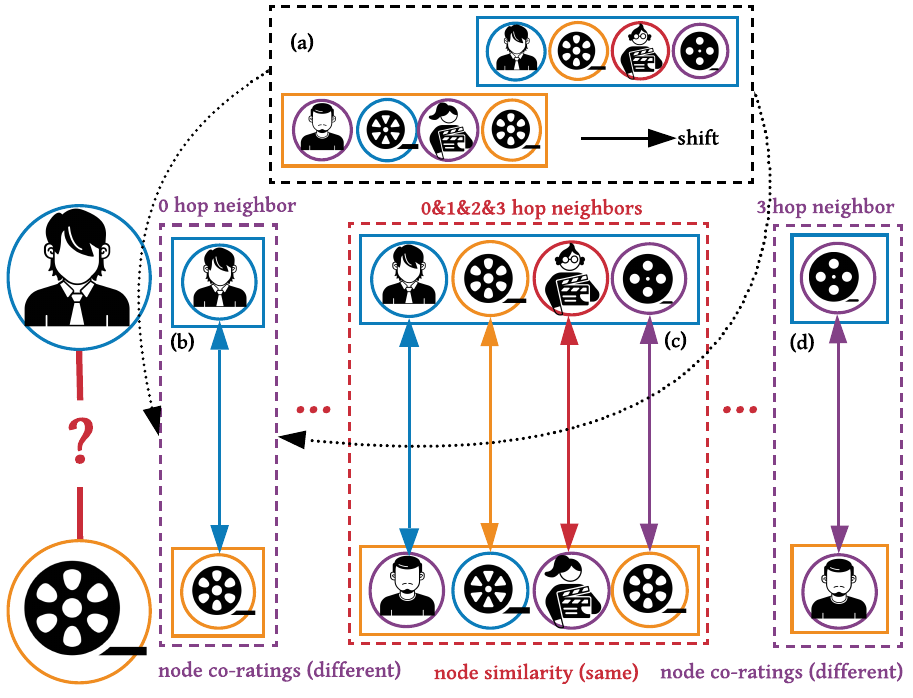}
	\vspace{-3mm}
	\caption {An illustrated example for the result after convolutional interaction operation. The result contains information of both node similarity (c) and node co-ratings (b) \& (d).}
	\label{fig:fft}
	
\end{figure}

Inspired by signal processing \cite{oppenheim1999discrete}, this neighborhood division and inner group interaction can be formulated as a unified operation named convolution.
Roughly speaking, the convolution mainly contains three kinds of operations, namely shift, product, and sum, which are employed repeatedly until they meet the end of the path.
Take Figure~\ref{fig:fft} for example. Our task is to calculate the interaction between source user neighborhood $u_\text{A}, m_\text{B}, d_\text{B}, m_\text{D}$ and target movie neighborhood $m_\text{B}, d_\text{C}, m_\text{C}, u_\text{C}$.
First, we inverse the order of target movie neighborhood and obtain $u_\text{C}, m_\text{C}, d_\text{C}, m_\text{B}$.
We shift it from left to right and observe the overlapping nodes during the shift.
As shown in Figure~\ref{fig:fft}(a), the first overlapping happens between source and target nodes, namely 0-hop neighbor.
We utilize the product operation and obtain the co-ratings between different types of nodes $r(u_\text{A}, m_\text{B})$ (as Figure~\ref{fig:fft}(b) shows).
Then, we repeatedly shift, product, and sum, and then reach the situation where all nodes are overlapped.
The result in this situation is the similarity between the same type of nodes $r(u_\text{A}, u_\text{C}) + r(m_\text{B}, m_\text{C}) + r(d_\text{B}, d_\text{C}) + r(m_\text{D}, m_\text{B})$ (as shown in Figure~\ref{fig:fft}(c)).
In a similar way, the last interaction happens between different types of nodes $r(m_\text{D}, u_\text{C})$ (as shown in Figure~\ref{fig:fft}(d)).

\begin{figure}[b]
	\centering
	\vspace{-3mm}
	\includegraphics[width=0.35\textwidth]{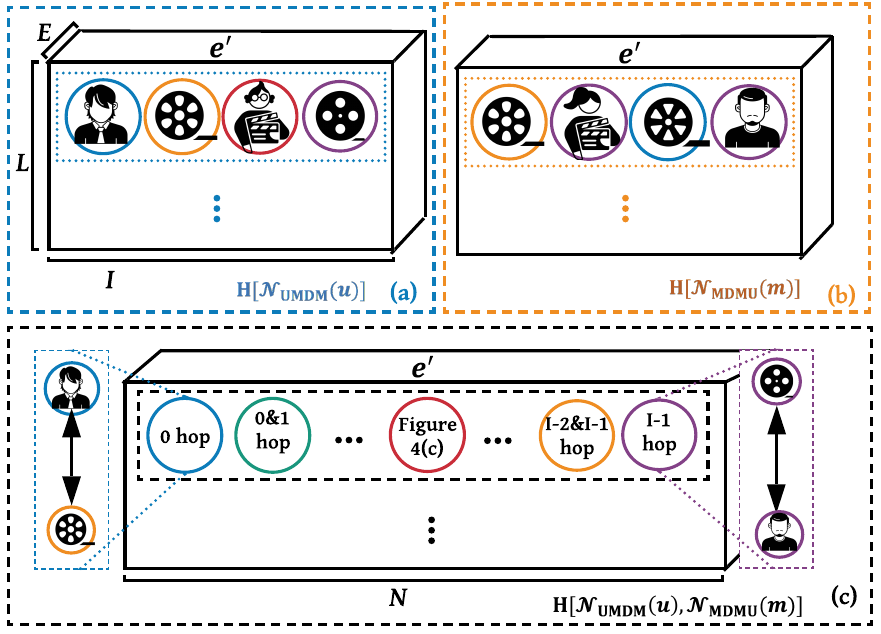}
	\vspace{-3mm}
	\caption {An illustrated example of the embedding matrix of metapath $\rho$ guided neighborhood of source node $s$ ($\mathbf{H}[\mathcal{N}_\rho(s)]$), and target node $t$ ($\mathbf{H}[\mathcal{N}_\rho(t)]$) generated according to Eq.~\ref{eqn:embedding}; and interaction matrix ($\mathbf{H}[\mathcal{N}_\rho(s), \mathcal{N}_\rho(t)]$) calculated according to Eq.~\ref{eqn:interact}.}
	\label{fig:cube}
	\vspace{-3mm}
\end{figure}

Let $\mathbf{H}[\mathcal{N}_\rho(o)]$ denote the embedding matrix of metapath $\rho$ guided neighbors of object $o$, which can be formulated as
\begin{equation}
	\label{eqn:embedding}
	\mathbf{H}[\mathcal{N}_\rho(o)]_l = [e^\rho_0 \oplus e^\rho_1 \oplus \cdots \oplus e^\rho_{I-1}],
\end{equation}
where $l$ represents the $l$-th metapath, $e_i^\rho$ means the embedding of the node in the $i$-th step of one metapath, $\oplus$ denotes the stack operation, $I$ means the metapath length.
Hence, as illustrated in Figure~\ref{fig:cube}, $\mathbf{H}[\mathcal{N}_\rho(o)]$ is a $\mathbb{R}^{L \times I \times E}$ matrix, where $L$ is the number of metapaths, $I$ is the length of the metapath, $E$ means the dimension of the node embedding.
Based on convolutional operations, we further define the interaction between neighborhoods of source and target objects as

\begin{equation}
	\mathbf{H}[\mathcal{N}_\rho(s), \mathcal{N}_\rho(t)]_{l} = \mathbf{H}[\mathcal{N}_\rho(s)]_{l} \odot \mathbf{H}[\mathcal{N}_\rho(t)]_{l},
\end{equation}
where $\odot$ denotes the convolutional operation.
According to the definition of convolution, one can write formulation as
\begin{equation}
	\label{eqn:interact}
	\mathbf{H}[\mathcal{N}_\rho(s), \mathcal{N}_\rho(t)]_{l, n} = \sum_{\substack{a, b\\ a+b ~ \text{mod} ~ N = n}} \mathbf{H}[\mathcal{N}_\rho(s)]_{l, a} \cdot \mathbf{H}[\mathcal{N}_\rho(t)]_{l, b}.
\end{equation}

One can find that $\mathbf{H}[\mathcal{N}_\rho(s), \mathcal{N}_\rho(t)] \in \mathbb{R}^{L \times N \times E}$, $N$ is the length of convolution outputs and it equals to $I_s + I_t - 1$ where $I_s$, $I_t$ denote the metapath length of source and that of target nodes respectively.

The well-known convolution theorem states that convolution operations in the spatial domain are equivalent to pointwise products in the Fourier domain. 
Let $\mathcal{F}$ denote the fast Fourier transformation (FFT) and $\mathcal{F}^{-1}$ its inverse, we can compute convolutions as
\begin{equation}
	\label{eqn:interaction}
	\mathbf{H}[\mathcal{N}_\rho(s), \mathcal{N}_\rho(t)] = \mathcal{F}^{-1}(\mathcal{F}(\mathbf{H}[\mathcal{N}_\rho(s)]) \cdot \mathcal{F}(\mathbf{H}[\mathcal{N}_\rho(t)])).
\end{equation}

Let $\mathbf{H}[\mathcal{N}_\rho]$ denote $\mathbf{H}[\mathcal{N}_\rho(s), \mathcal{N}_\rho(t)]$ in the following sections for convenience.
As stated in \cite{mathieu2013fast}, the time complexity of plain convolution is $O(I^2)$, and it is reduced to $O(I \log (I))$ when using FFT.	
According to the analysis above, we can conclude that not only can this structure capture both node similarity and co-ratings in grouped neighborhood, but also it can implement with high efficiency.

\subsection{Aggregation Module}
\label{subsec:aggregation}
In this section, we consider the aggregation module in two sides.
On the first side, from Figure~\ref{fig:cube}, we can see that elements in interaction matrix $\mathbf{H}[\mathcal{N}_\rho] = \mathbf{H}[\mathcal{N}_\rho(s), \mathcal{N}_\rho(t)]$, contain interactions between various types of nodes.
Hence, it is natural to capture the key interaction in element/node-level during aggregation procedure.
On the other side, every object $o$ on HIN contains multiple types of semantic information represented with different metapath $\rho_0, \rho_1, \cdots, \rho_{P-1}$.
This further causes various interaction matrix $\mathbf{H}[\mathcal{N}_{\rho_0}]$, $\mathbf{H}[\mathcal{N}_{\rho_1}]$, $\cdots$, $\mathbf{H}[\mathcal{N}_{\rho_{P-1}}]$.
To capture key message in a complex graph, we need to fuse multiple semantics revealed by different metapath, \emph{i.e.}, in path/matrix-level. 

\minisection{Node/Element-level Attention} 
Similar as \cite{wang2019heterogeneous}, we leverage a self-attention mechanism to learn the weights among various kinds of nodes in metapath $\rho$ as
\begin{equation}
	h^\rho_{ij} = (h^\rho_{i}W_T) \cdot (h^\rho_{j}W_S),
\end{equation}
where $h^\rho_{i}$, $h^\rho_{j}$ are elements of interaction matrix $\mathbf{H}[\mathcal{N}_\rho]$ and $W_T, W_S$ are trainable weights.
$h^\rho_{ij}$ can show how important interaction $h_j$ will be for interaction $h_i$.
To retrieve a general attention value in metapath $\rho$, we further normalize this value in neighborhood scope as
\begin{equation}
	\alpha^{\rho}_{ij} = \text{softmax}(h^\rho_{ij}) = \frac{\text{exp}(h^\rho_{ij}/\iota)}{\sum_{j\in \mathcal{N_\rho}} \text{exp}(h^\rho_{ij}/\iota)},
\end{equation}
where $\iota$ denotes the temperature factor, and $\mathcal{N}_\rho$ here covers all the interaction elements/nodes guided by $\rho$, namely the entirely $\mathbf{H}[\mathcal{N}_\rho]$.
To jointly attend to the neighborhoods from different representation subspaces and learn stably, we leverage the multi-head attention as in previous works \cite{wang2019heterogeneous} to extend the observation as
\begin{equation}
	\label{eqn:node_attention}
	z^\rho_i = \sigma(W_q \cdot (\frac{1}{H} \sum_{n=1}^H \sum_{j\in \mathcal{N}_\rho} \alpha^{\rho}_{ijn}(h^\rho_j W_C^n))+b_q),
\end{equation}
where $H$ is the number of attention heads, and $W_q, W_C, b_q$ are trainable parameters.
Hence, the metapath based embedding $z^\rho_i$ is aggregated based on metapath-guided neighborhood with single metapath and semantic-specific information, \emph{i.e.}, $\mathbf{H}[\mathcal{N}_\rho]$.
Given the metapath set $\{\rho_0, \rho_1, \cdots, \rho_{P-1}\}$, after feeding into node-level attention, we can obtain $P$ groups of semantic-specific interaction embeddings, denotes as $\{\mathbf{Z}[\mathcal{N}_{\rho_0}], \mathbf{Z}[\mathcal{N}_{\rho_1}], \cdots,  \mathbf{Z}[\mathcal{N}_{\rho_{P-1}}]\}$.\par

\minisection{Path/Matrix-level Attention}
To obtain the importance of each metapath, we first transform the semantic-specific embedding through a nonlinear transformation.
We then measure the path-level attention value as the average of the importance of all the semantic-specific node-level embeddings.
The importance of each metapath $\rho_j$, denoted as $\omega^{\rho_j}$, is shown as follows:
\begin{equation}
	\omega^{\rho_j} = \frac{1}{|\mathcal{V}|} \sum_{i \in \mathcal{V}} w^T \cdot \text{tanh} (W_q \cdot z^{\rho_j}_i + b_q),
\end{equation}
where $w$ is a path-level attention vector.
We then normalize the above importance of all metapaths via a softmax function
\begin{equation}
	\beta^{\rho_j} = \text{softmax}(\omega^{\rho_j}) = \frac{\text{exp}(\omega^{\rho_j}/\tau)}{\sum_{j=0}^{P-1}\text{exp}(\omega^{\rho_j}/\tau)},
\end{equation}
where $\tau$ is the temperature factor.
It can be explained as the contribution of the metapath $\rho_j$ in a specific task.
With the learned weights as coefficients, we can fuse these semantic-specific embeddings to obtain the final embedding $Z$ via
\begin{equation}
	\label{eqn:semantic_attention}
	Z = \sum_{j=0}^{P-1} \beta^{\rho_j} \cdot \mathbf{Z}[\mathcal{N}_{\rho_j}].
\end{equation}
Hence, we have obtained the final aggregation result, which involves interaction information in both node- and path-level.

\subsection{Optimization Objective}
The final prediction result $\hat{Y}$ can be derived from final embedding $Z$ via a nonlinear projection (\emph{e.g.}, MLP).
The loss function of our model is a log loss:
\begin{equation}
	\label{eqn:loss}
	\mathcal{L}(Y, \hat{Y}) = \sum_{i, j \in \mathcal{Y}^+ \bigcup \mathcal{Y}^-} (y_{ij} \log \hat{y}_{ij} + (1- y_{ij}) \log(1-\hat{y}_{ij}))
\end{equation}
where $y_{ij}$ is the label of the instance and  $\mathcal{Y}^+$, $\mathcal{Y}^-$ denote the postive instances set and the negative instances set, respectively.

\begin{table*}
	\centering
	\caption{The results of CTR prediction in terms of AUC, ACC. \emph{Note}: ``*'' indicates the statistically significant improvements over the best baseline, with $p$-value smaller than $10^{-6}$ in two-sided $t$-test.}
	\label{tab:ctr}
	\vspace{-3mm}
	\begin{tabular}{|c|cc|cc|cc|cc|}
		\hline
		\multirow{2}{*}{Model} & \multicolumn{2}{c|}{Movielens} & \multicolumn{2}{c|}{LastFM} & \multicolumn{2}{c|}{AMiner} & \multicolumn{2}{c|}{Amazon} \\ 
		\cline{2-9}
		& AUC & ACC & AUC & ACC & AUC & ACC & AUC & ACC\\ \hline
		NeuMF \cite{he2017neural} & 0.7890 & 0.7378 & 0.8900 & 0.8102 & 0.8130 & 0.7897& 0.6841 & 0.6405\\
		\hline
		HAN \cite{wang2019heterogeneous} & 0.8110 & 0.7530 & 0.9113 &	0.8289 & 0.8451 & 0.8284 & 0.7207 &	0.6831\\ 
		\hline
		HetGNN \cite{zhang2019heterogeneous} & 0.7830 & 0.7411 & 0.9020 & 0.8270 & 0.8202 & 0.7939 & 0.7061 & 0.6627 \\
		\hline
		LGRec \cite{hu2018local} & 0.8030 & 0.7504 & 0.9127 & 0.8331 & 0.8308 & 0.8130 & 0.7058 & 0.6572 \\ 
		\hline
		MCRec \cite{hu2018leveraging} & 0.8161 & 0.7622 & 0.9274 & 0.8471 & 0.8512 & 0.8339 & 0.7274 & 0.6940 \\
		\hline
		IPE \cite{liu2018interactive} & 0.8186 & 0.7693 & 0.9235 & 0.8440 & 0.8411 & 0.8209 & 0.7173 & 0.6789 \\
		\hline
		\hline
		NIRec$_\text{CNN}$ & 0.8342 & 0.7777 & 0.9353 & 0.8593 & 0.8734 & 0.8504 & 0.7397 & 0.7060\\
		\hline
		NIRec$_\text{GCN}$ & 0.8290 & 0.7630 & 0.9290 & 0.8571 & 0.8636 & 0.8475 & 0.7379 & 0.7042 \\ 
		\hline
		NIRec & \textbf{0.8468}$^*$ & \textbf{0.7896}$^*$ & \textbf{0.9404}$^*$ & \textbf{0.8665}$^*$ & \textbf{0.8760}$^*$ & \textbf{0.8562}$^*$ & \textbf{0.7493}$^*$ & \textbf{0.7110}$^*$\\
		\hline
	\end{tabular}
\vspace{-3mm}
\end{table*}

\vspace{-1mm}
\subsection{Model Analysis}
\label{sec:analysis}
The learning algorithm of NIRec is given in Section~\ref{apsec:algo} in the appendix. 
Here we give the analysis of the proposed NIRec as follows:

\minisection{Complexity}The proposed NIRec is highly efficient and can be easily parallelized. 
We provide the model efficiency analysis for both interaction module and aggregation module. 
As stated in \cite{mathieu2013fast}, in interaction module, the complexity of the FFT-based method requires $O(L^2_\rho I_\rho \log(I_\rho))$ where $L_\rho$ and $I_\rho$ denote the number of path guided by $\rho$ and metapath length respectively.
As for aggregation module, given a metapath $\rho$, the time complexity of node-level attention is $O(V_\rho K + E_\rho K)$, where $V_\rho$ is the number of nodes, $E_\rho$ is the number of metapath based node pairs, $K$ is the number of attention heads.
The computation of attention can compute individually across all nodes and metapaths.
The overall complexity is linear to the number of nodes and metapath based node pairs. The proposed model can be easily parallelized, because the node- and path-level attention can be parallelized across node pairs and metapaths, respectively. The overall complexity is linear to the number of nodes and metapath based node pairs.

\minisection{Interpretability}
The proposed model has potentially good interpretability for the learned interaction embedding through similarity and co-ratings among neighbor nodes.
Also, with the learned importance in node- and path-level, the proposed model can pay more attention to some meaningful interactions or metapaths for the specific task and give a more comprehensive description of a heterogeneous graph.
Based on the attention values, we can check which interactions or metapaths make the higher (or lower) contributions for our task, which is beneficial to analyze and explain our results.

\vspace{-2mm}

\section{Experiments}
\label{sec:exp}
In this section, we present the details of the experiment setups and the corresponding results. 
To illustrate the effectiveness of our proposed model, we compare it with some strong baselines on recommendation task.
We start with three research questions (RQ) to lead the experiments and the following discussions.
\begin{itemize}[topsep = 3pt,leftmargin =5pt]
	\item (\textbf{RQ1}) Compared with the baseline models, does NIRec achieve state-of-the-art performance in recommendation tasks on HINs?
	\item (\textbf{RQ2}) What is the influence of different components in NIRec? Are the proposed interaction and aggregation modules necessary for improving performance?
	\item (\textbf{RQ3}) What patterns does the proposed model capture for the final
	recommendation decision?
	\item (\textbf{RQ4}) How do various hyper-parameters, \emph{i.e.}, the length and type of metapath-guided neighborhood, impact the model performance?
\end{itemize}
\vspace{-3mm}

\subsection{Dataset Description}
\label{apsec:data}
We adopt four widely used datasets from different domains, namely Movielens movie dataset\footnote{\url{https://grouplens.org/datasets/movielens/}}, LastFM music dataset\footnote{\url{https://grouplens.org/datasets/hetrec-2011/}}, AMiner paper dataset\footnote{\url{https://AMiner.org/data}} and Amazon e-commerce dataset\footnote{\url{http://jmcauley.ucsd.edu/data/amazon/}}.
We treat a rating as an interaction record, indicating whether a user has rated an item.
Also, we provide the main statistics of four datasets are summarized in Table~\ref{tab:data} in appendix to help with reproducibility.
The first row of each dataset corresponds to the numbers of users, items and interactions, while the other rows correspond to the statistics of other relations. 
We also report the selected metapaths for each dataset in the last column of the table.
The detailed data preprocessing of is given at Section~\ref{apsec:data} in appendix.
\vspace{-2mm}

\subsection{Compared Methods}
\label{subsection:compare}
We use six baseline methods containing heterogeneous and attributed graph embedding models such as HAN, HetGNN, and IPE, as well as recommendation models such as NeuMF, LGRec, and MCRec.
It is worth noting that, HAN, HetGNN, IPE, and MCRec are recently proposed, state-of-the-art models.
\begin{itemize}[topsep = 3pt,leftmargin =5pt]
	\item \textbf{NeuMF}: \citet{he2017neural} introduced a generalized model consisting of a matrix factorization (MF) component and an MLP component.
	\item \textbf{HAN}: \citet{wang2019heterogeneous} introduced hierarchical attention to capture node-level and semantic-level information.
	\item \textbf{HetGNN}:  \citet{zhang2019heterogeneous} introduced an unified framework to jointly consider heterogeneous structural information as well as heterogeneous contents information, adaptive to various HIN tasks.
	\item \textbf{LGRec}: \citet{hu2018local} proposed a unified model to explore and fuse local and global information for recommendation.
	\item \textbf{MCRec}: \citet{hu2018leveraging} leveraged rich metapath-based context to enhance the recommendation performance on HINs. 
	\item \textbf{IPE}: \citet{liu2018interactive} proposed interactive paths embedding to capture rich interaction information among metapaths.		
\end{itemize}

In order to investigate the impact of different components in our model, we set several variants of NIRec model as baselines.

\begin{itemize}[topsep = 3pt,leftmargin =5pt]
	\item \textbf{NIRec$_\text{CNN}$}: a variant of NIRec model which employs Convolutional Neural Networks (CNN) to capture contextual information within the source/target neighborhood without any interaction.
	\item \textbf{NIRec$_\text{GCN}$}: is another variant of NIRec model which employs Graph Convolutional Networks (GCN) to aggregate interaction information instead of attention mechanism adopted in this paper.
\end{itemize}
One can observe that \textbf{NIRec$_\text{CNN}$}, \textbf{NIRec$_\text{GCN}$} are designed to test performance gains from interaction and aggregation module, respectively.
We also provide detailed configuration in Section~\ref{apsec:variant} in the appendix.
\vspace{-3mm}

\subsection{Result Analysis}
\label{subsec:analysis}
We evaluate these models on the click-through rate prediction (CTR) task.
We use the metrics Area Under Curve (AUC) and Accuracy (ACC), which are widely used in binary classification problems. The details of the implementation is given in Section~\ref{apsec:detail}.

\minisection{Experimental Results and Analysis \textbf{(RQ1)}}
\label{minisec:analysis}
The comparison results of our proposed model and baselines on four datasets are reported in Table~\ref{tab:ctr}.
The major findings from experimental results are summarized as follows:
\begin{itemize}[topsep = 3pt,leftmargin =5pt]
	\item Our model NIRec is consistently better than all baselines on the four datasets.
	The results indicate the effectiveness of NIRec on the task of CTR prediction, which has adopted a principled way to leverage interaction information for improving recommendation performance.
	\item Among the two kinds of baselines, most metapath-based methods (HAN, MCRec, IPE) outperform graph-based or feature-based methods (HetGNN, NeuMF) in most cases.
	An intuitive explanation is that those metapath-based methods can better capture the rich high-order structural information on HINs.
	It should be noted that our model NIRec based on metapath-guided neighborhood is able to jointly consider first-order neighbor information with high-order semantic message.
	\item Among HIN-based baselines, the recently proposed methods MCRec and IPE gain better performance than the others.
	It is easy to notice that both of them try to capture context information or interactive patterns among paths.
	A possible reason is that simple aggregation of semantic message on paths may lose some key information.
	It should be noted that NIRec not only captures interaction information but also has potentially good interpretability.
\end{itemize}

\minisection{Ablation Study \textbf{(RQ2)}}
In order to investigate the contribution of each component to the final recommendation performance, we design two variants of NIRec, namely NIRec$_\text{CNN}$ and NIRec$_\text{GCN}$, to study interaction and aggregation modules, respectively.
The results are shown in Table~\ref{tab:ctr}.
The findings are in two aspects. 
First, it (NIRec > NIRec$_\text{CNN}$) indicates that our convolutional interaction strategy is able to better capture interaction information (\emph{i.e.}, similarities between the same type of nodes and ratings between different types of nodes) than simply employing CNN layers.
Also, it (NIRec > NIRec$_\text{GCN}$) shows that the attention mechanism can better utilize the metapath-based interactive information.
The node-level interactions and path-level metapaths may contribute differently to the final performance.
Ignoring such influence may not be able to achieve optimal performance. 

\begin{figure}[h]
	\vspace{-3mm}
	\centering
	\includegraphics[width=0.45\textwidth]{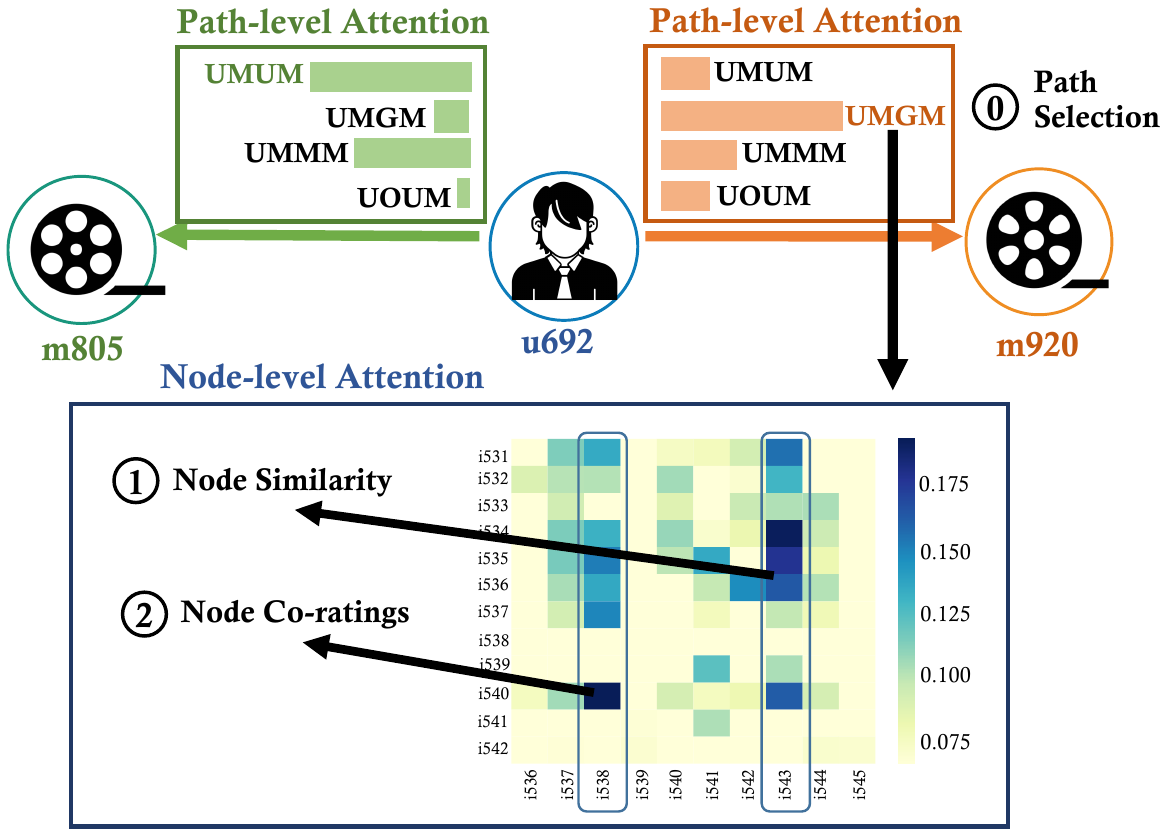}
	\vspace{-3mm}
	\caption {An illustrative example of the interpretability of interaction-specific attention distributions for NIRec. The number denotes the logic flow of interpretation.}
	\label{fig:case}
	\vspace{-3mm}
\end{figure}

\begin{figure*}[t]
	\centering
	\includegraphics[width=1\textwidth]{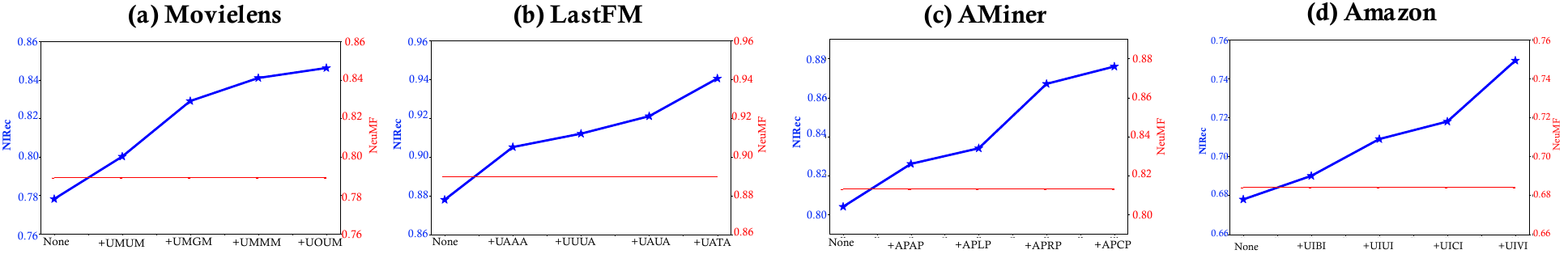}
	\vspace{-7mm}
	\caption {Performance change of NIRec when gradually incorporating metapaths in terms of AUC.}
	\label{fig:path}
	\vspace{-3mm}
\end{figure*}

\minisection{Case Study \textbf{(RQ3)}}
A major contribution of NIRec is the incorporation of the node- and path-level attention mechanism, which takes the interaction relation into consideration in learning effective representations for recommendation.
Besides the performance effectiveness, another benefit of the attention mechanism is that it makes the recommendation result highly interpretable.
To see this, we select the user u692 in the Movielens dataset as an illustrative example.
Two interaction records of this user have been used here, namely m805 and m920.
In Figure~\ref{fig:case}, we can see that each user-movie pair corresponds to a unique attention distribution, summarizing the contributions of the metapaths. 
The relation between u692-m805 pair mainly relies on the metapaths UMUM and UMMM, while u692-m920 pair mainly relies on metapath UMGM (denoted by \circled{0} in Figure~\ref{fig:case}). 
By inspecting into the dataset, it is found that at least five first-order neighbors of u692 have watched m805, which explains why user-oriented metapaths UMUM and UUUM play the key role in the first pair. 
As for the second pair, we find that the genre of m805 is g3, which is the favorite movie genre of u692. 
This explains why genre-oriented metapath UMGM plays
the key role in the second interaction. 
Our path-level attention is able to produce path-specific attention distributions.\par

If we wonder more specific reasons, we can have a look at the  node-level attention.
Here, we plot the interactive attention values in Figure~\ref{fig:case}.
We can observe that the attention value of the similarity for the same type nodes is very high (denoted by \circled{1} in Figure~\ref{fig:case}).
By inspecting into the dataset, we can find that there are metapaths connected between u692 and m920.
In other words, some parts of the metapath-guided neighborhood of u692 and m920 overlap each other, which causes high similarity.
Also, the co-ratings between two neighborhoods play another key role (denoted by \circled{2} in Figure~\ref{fig:case}).
It is natural, since in these neighborhoods, many users are fans of g3 and movies are in type of g3, which causes the co-ratings among them becoming really high. 
According to the analysis above, we can see that the distributions of attention weights are indeed very skew, indicating some interactions and metapaths are more important to consider than the others.

\begin{figure}[t]
	\centering
	\includegraphics[width=0.5\textwidth]{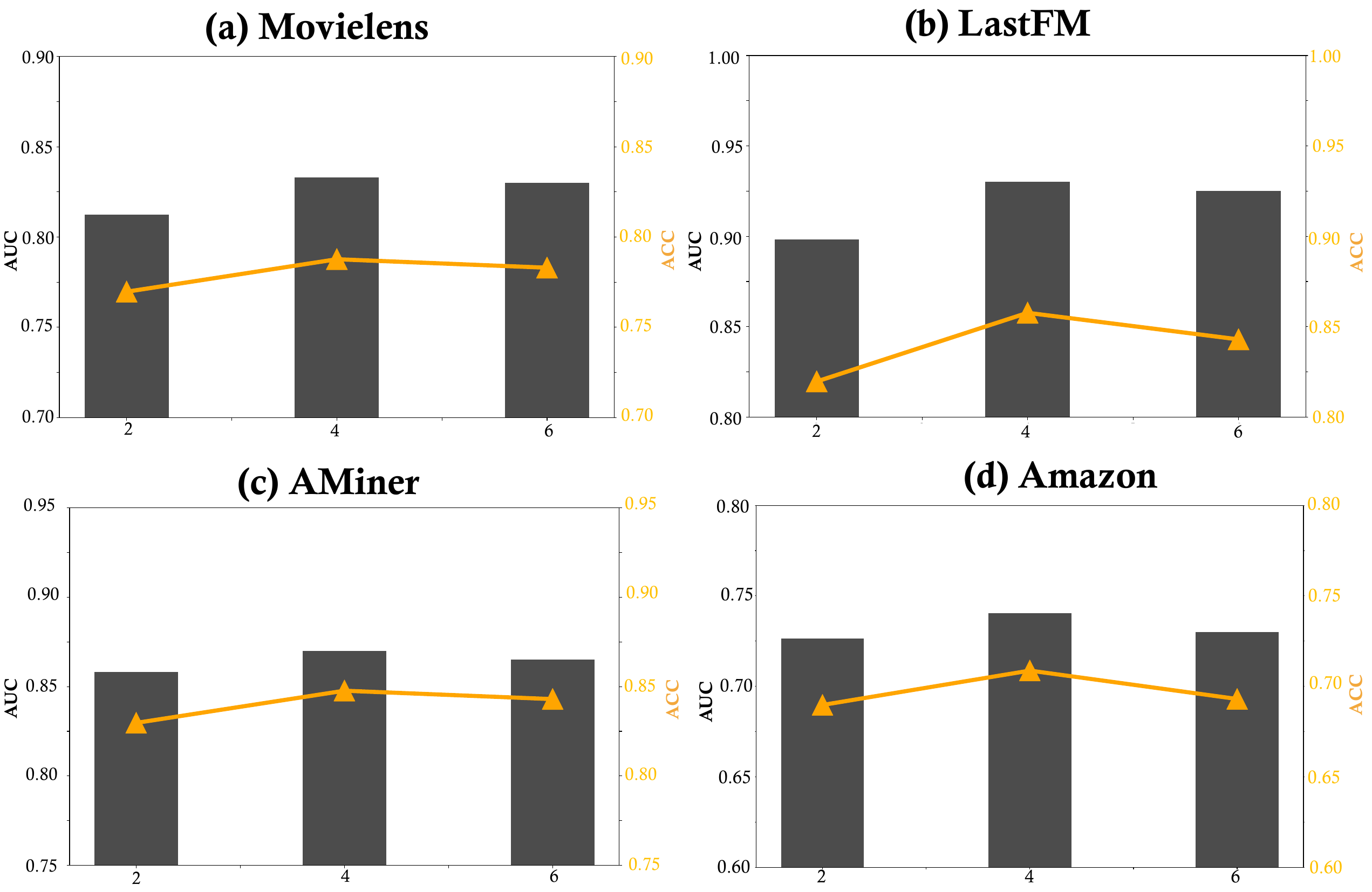}
	\vspace{-7mm}
	\caption {Performance change of NIRec with different neighborhood length in terms of AUC and ACC.}
	\label{fig:neighbor}
	\vspace{-2mm}
\end{figure}

\minisection{Impact of Metapath \textbf{(RQ4)}}
In this section, we investigate the impact of different metapaths on the recommendation performance through gradually incorporating metapaths into the proposed model. 
For ease of analysis, we include the NeuMF as the reference baseline. 
In Figure~\ref{fig:path}, we can observe that the performance
of NIRec overall improves with the incorporation of more metapaths. 
Meanwhile, metapaths seem to have different effects on the recommendation performance. 
Particularly, we can find that, when adding UMGM, NIRec has a significant performance boost in the Movielens dataset (Figure~\ref{fig:path}(a)).
Similar situation happens when adding UIVI in the Amazon dataset (Figure~\ref{fig:path}(b)). 
These findings indicate that different metapaths contribute differently in the final result, consistent with previous observations in Section~\ref{minisec:analysis},

\minisection{Impact of N-Hop Neighborhood \textbf{(RQ4)}}
In this section, we investigate the impact of different lengths of neighborhoods, \emph{i.e.}, n-hop neighborhood.
When changing the length of the neighborhood, we make other factors, \emph{e.g.}, metapath type fixed.
For example, when the metapath is UMGM, we study UM for length 2, UMGM for length 4, and UMGMGM for length 6.
We conduct similar procedures for the other metapaths and obtain the results in Figure~\ref{fig:neighbor}.
We can observe that the performance first improves and then declines when the length of the neighborhood increases.
Particularly, we can find that NIRec reaches the best performance in the metapath with length 4.
The possible reason may be that as the length of the neighborhood increases, the metapath-guided neighborhood can maintain more information.
When the length of the neighborhood is smaller than 4, the information is mainly useful for the final performance.
However, when the length exceeds 4, the information includes noisy message which harms the recommendation performance.
These findings indicate that different lengths of metapaths contribute differently to the final performance.
Also, it should be noted that our model is able to interact and aggregate neighbors in different lengths, as illustrated in Figure~\ref{fig:overview}.

\section{Conclusion and Future Work}
\label{sec:conclusion}
In this paper, we introduced the problem of ``early summarization'' and proposed a neighborhood-based interaction-enhanced recommendation model, \emph{i.e.}, NIRec, to address this problem.
We first introduce the definition of metapath-guided neighborhoods to preserve the heterogeneity on HINs.
Then, we elaborately designed an interaction module to capture the similarity of each source and target node pair through their neighborhoods.
To fuse the rich semantic information, we proposed the node- and path-level attention mechanism to capture the key interaction and metapath, respectively. 
Extensive experimental results have demonstrated the superiority of our model in both recommendation effectiveness and interpretability.
Currently, our approach is able to capture interactive information only in the structural (graph) side effectively.
However, there is rich semantic information on both the structural (graph) and non-structural (node) sides.
In the future, a promising direction is extending neighborhood interaction and aggregation modules to capture key message from two sides and adapt to more general scenarios.

\minisection{Acknowledgments} 
The corresponding author Weinan Zhang thanks the support of National Natural Science Foundation of China (Grant No. 61702327, 61772333, 61632017).
We would also like to thank Wu Wen Jun Honorary Doctoral Scholarship from AI Institute, Shanghai Jiao Tong University.

\bibliographystyle{ACM-Reference-Format}
\balance
\bibliography{hinrec}

\clearpage
\appendix
\section{appendix}
\subsection{Pseudocode of NIRec Training Procedure}
\label{apsec:algo}
The pseudocode of the NIRec training procedure is described in Algorithm~\ref{algo:framework}. 
In the HIN-based recommendation scenario, each source node $n_s$ corresponds to a user $u_s$, and each target node $n_t$ corresponds to an item $i_t$.
The task in this work can be either stated as link predictions on HINs or CTR predictions on HIN-based recommendation.

\begin{algorithm}[!h]
	\caption{NIRec}
	\label{algo:framework}
	\begin{algorithmic}[1]
		\REQUIRE
		HIN $\mathcal{G} = (\mathcal{V}, \mathcal{E})$; node feature $\{e_, i \in \mathcal{V}\}$; metapath set $\{\rho_0, \rho_1, \cdots, \rho_{P-1}\}$; source code $n_s$ and target node $n_t$
		\ENSURE
		final link prediction $\hat{Y}$ between $n_s$ and $n_t$
		\vspace{1mm}
		\STATE Initialize all parameters.
		\REPEAT 
		\FOR {each metapath $\rho_k \in \{\rho_0, \rho_1, \cdots \rho_{P-1}\}$}
		\STATE Find metapath-guided neighborhoods of $n_s$, $n_t$: $\mathcal{N}_{\rho_k}(n_s)$, $\mathcal{N}_{\rho_k}(n_t)$ according to Eq.~\ref{eqn:neighbor}.\\
		\STATE Obtain interaction result $\mathbf{H}[\mathcal{N}_{\rho_k}(n_s), \mathcal{N}_{\rho_k}(n_t)]$ according to Eq~\ref{eqn:interaction}. \\
		\STATE Calculate node/element-level embedding $z^{\rho_k}$ according to Eq.~\ref{eqn:node_attention}.
		\ENDFOR
		\STATE Fuse path/matrix-level embedding $Z$ according to Eq.~\ref{eqn:semantic_attention}.\\
		\STATE Obtain final predication $\hat{Y}$ via MLP.\\
		\STATE Calculate loss $\mathcal{L}(Y, \hat{Y})$ according to Eq.~\ref{eqn:loss}, and Back propagation.\\
		\UNTIL convergence
	\end{algorithmic}
\end{algorithm}

As Algorithm~\ref{algo:framework} shows, we first generate neighborhoods $\mathcal{N}_{\rho_k}(n_s)$, $\mathcal{N}_{\rho_k}(n_t)$ of $n_s$, $n_t$ guided by metapath $\rho_k$ in line 4.
We then obtain interactive information $\mathbf{H}[\mathcal{N}_{\rho_k}(n_s), \mathcal{N}_{\rho_k}(n_t)]$ through the interaction module described in Section~\ref{subsec:interaction} in line 5.
To fuse the interaction results, we leverage the node-level attention mechanism to obtain representation $z^{\rho_k}$ in line 6.
We repeat the above procedure for different metapaths and obtain various node-level embeddings $z^{\rho_0}, z^{\rho_1}, \cdots, z^{\rho_{P-1}}$.
To capture the key metapath in the current task, we employ the path-level attention mechanism to get embedding $Z$ in line 8.
Finally, we obtain the final predication $\hat{Y}$ via MLP layers in line 9.

\subsection{Data Preproccess}	
Each dataset is processed as follows.
Since the relations between users and items are originally in rating format, we convert ratings to binary feedbacks: ratings with 4--5 stars are converted to positive feedbacks (denoted as ``1''), and other ratings are converted to negative feedbacks (denoted as ``0'').
After the datasets are processed, we split each dataset into training/validation/test sets at a ratio of 6:2:2.

\subsection{Model Variants Configuration}
\label{apsec:variant}

\begin{itemize}[topsep = 3pt,leftmargin =5pt]
	\item \textbf{NIRec$_\text{CNN}$}: This variant does not consider the interpretability in the interaction module.
	That is, it directly fuses the neighborhood of each source and target node via Convolutional Neural Networks (CNN).
	This context embedding method is quite similar with the path instance embedding technique in \cite{hu2018leveraging}.
	The aggregation module is the same as NIRec.
	Hence, this variant is designed to show the performance gain by our interaction module.
	\item \textbf{NIRec$_\text{GCN}$}: This variant replaces the attention mechanism of NIRec with a standard Graph Convolutional Network (GCN).
	That is, both the node- and path-level features are fed to GCN layers to get content embeddings without fully considering different types of nodes and metapaths on HINs.
	The interaction module is the same as NIRec.
	Hence, this variant is designed to show the performance gain by our aggregation module.
\end{itemize}

\begin{table}[!h]\footnotesize
	\centering
	\caption{Statistics of the four datasets. The last column reports the selected metapaths in each dataset.}
	\vspace{-3mm}
	\label{tab:data}
	\begin{tabular}{|c|c|c|c|c|c|}
		\hline
		Datasets & Relations (A-B) & A & B & A-B & Metapath \\
		\hline
		\multirow{4}{*}{Movielens} & User-Movie & 943 & 1,682 & 100,000 & UMUM \\
		\cline{2-6}
		& Movie-Movie & 1,682 & 1,682 & 82,798 & UMMM \\
		\cline{2-6}
		& User-Occupation & 943 & 21 & 943 & UOUM \\
		\cline{2-6}
		& Movie-Genre & 1,682 & 18 & 2,861 & UMGM \\ 
		\cline{2-6}
		\hline
		\multirow{4}{*}{LastFM} & User-Artist & 1,892 & 17,632 & 92,834 & UAUA \\
		\cline{2-6}
		& User-User & 1,892 & 1,892 & 18,802 & UUUA \\
		\cline{2-6}
		& Artist-Artist & 17,632 & 17,632 & 153,399 & UAAA \\
		\cline{2-6}
		& Artist-Tag & 17,632 & 11,945 & 184,941 & UATA \\ 
		\cline{2-6}
		\hline
		\multirow{4}{*}{AMiner} & Author-Paper & 164,472 & 127,623 & 355,072 & APAP \\
		\cline{2-6}
		& Paper-Conference & 127,623 & 101 & 127,632 & APCP \\
		\cline{2-6}
		& Paper-Label & 127,623 & 10 & 127,623 & APLP \\
		\cline{2-6}
		& Paper-Reference & 127,623 & 147,251 & 392,519 & APRP \\ 
		\hline
		\multirow{4}{*}{Amazon} & User-Item & 3,584 & 2,753 & 50,903 & UIUI \\
		\cline{2-6}
		& Item-View & 2,753 & 3857 & 5,694 & UIVI \\
		\cline{2-6}
		& Item-Brand & 2,753 & 334 & 2,753 & UIBI \\
		\cline{2-6}
		& Item-Category & 2,753 & 22 & 5,508 & UICI \\ 
		\hline
	\end{tabular}
	\vspace{-3mm}
\end{table}

\subsection{Implementation Details}
\label{apsec:detail}

\minisection{Hyper-parameters}
The embedding dimension of NIRec is set to 128.
As stated in Section~\ref{subsec:sample}, we perform the metapath-guided random walk method. 
The size of sampled neighbors set of each node equals to 20 guided by the current metapath.
The length of metapath-guided neighborhood equals to the length of the metapath.
We investigate the impacts of different lengths of neighborhoods, \emph{i.e.}, n-hop neighborhoods in Section~\ref{subsec:analysis}.

\minisection{Baseline settings}
For fair comparisons, we employ the proposed method and baselines as following:
(i) In the data processing procedure, we sample and store paths guided by metapaths in a unified framework.
For instance, when sampling data from Movielens following UMUM-path, we can obtain 10 metapaths between $u_0$ and $m_0$, \emph{i.e.}, $\rho(u_0-m_0)$, as well as 10 metapath-guided neighborhoods of $u_0$, \emph{i.e.}, $\mathcal{N}(u_0)$, and 10 metapath-guided neighborhoods of $m_0$, \emph{i.e.}, $\mathcal{N}(m_0)$.
That is, the first node of both metapath $\rho(u_0-m_0)$ and metapath-guided neighborhood $\mathcal{N}(u_0)$ is the same source node $u_0$.
The difference is that the metapath must end with target node $m_0$ while the neighborhood can end with any node in the type of movie.
(ii) For metapath-based baselines such as MCRec and IPE, we feed models with the same paths.
For GNN-based baselines, such as HAN and HetGNN, we generate n-hop neighbor nodes from sampled paths.
(iii) The embedding dimensions of all baselines are set to 128 (same as NIRec).

\end{document}